# A machine learning framework for quantifying chemical segregation and microstructural features in atom probe tomography data

Alaukik Saxena[1], Nikita Polin[1], Navyanth Kusampudi[1], Shyam Katnagallu[1], Leopoldo Molina-Luna[4], Oliver Gutfleisch[5], Benjamin Berkels[3], Baptiste Gault[1,2], Jörg Neugebauer[1] and Christoph Freysoldt[1]

[1] *Max Planck Institut für Eisenforschung GmbH, Max Planck Strasse 1, 40237 Düsseldorf, Germany*

[2] *Department of Materials, Imperial College London, UK*

[3] *Aachen Institute for Advanced Study in Computational Engineering Science (AICES), RWTH Aachen University, Germany*

[4] *Advanced Electron Microscopy Division, Technische Universität Darmstadt, Darmstadt, Germany*

[5] *Functional Materials, Technische Universität Darmstadt, Darmstadt, Germany*

*Corresponding Author: Christoph Freysoldt <freysoldt@mpie.de>*

## Abstract

Atom probe tomography (APT) is ideally suited to characterize and understand the inter-play of chemical segregation and microstructure in modern multi-component materials. Yet, the quantitative analysis typically relies on human expertise to define regions of interest. We introduce a computationally efficient, multi-stage machine learning strategy to identify chemically distinct domains in a semi-automated way, and subsequently quantify their geometric and compositional characteristics. In our algorithmic pipeline, we first coarse-grain the APT data into voxels, collect the composition statistics, and decompose it via clustering in com-position space. The composition classification then enables the real-space segmentation via a density-based clustering algorithm, thus revealing the microstructure at voxel resolution. Our approach is demonstrated for a Sm-(Co,Fe)-Zr-Cu alloy. The alloy exhibits two precipitate phases with a plate-like, but intertwined morphology. The primary segmentation is further refined to disentangle these geometrically complex precipitates into individual plate-like parts by an unsupervised approach based on principle component analysis, or a U-Net-based semantic segmentation trained on the former. Following the chemical and geometric analysis, detailed chemical distribution and segregation effects relative to the predominant plate-like geometry can be readily mapped without resorting to the initial voxelization.

# Introduction

Atom probe tomography (APT) is a unique technique that provides the three-dimensional (3D) distribution of atoms in a material at sub-nanometer resolution along with their chemical identities Gault et al. (2012a, 2021); Lefebvre et al. (2016); Miller & Forbes (2014). APT provides insights into the composition of a variety of microstructural features discernible through their chemical fingerprint, including grain boundaries, dislocations, and secondary phase precipitates Prithiv et al. (2022); Zhou et al. (2021); Medrano et al. (2018). The interplay between these microstructural features along with their chemistry, on the other hand, determines the mate-rial's macroscopic properties. For exploring structure-property relations and, ultimately, for tailoring new materials, the individual objects forming the microstructure must be quantified in abundance, size, composition, geometrical shape, etc.

Obtaining such a comprehensive quantitative description from APT datasets in a reliable, controlled, and reproducible way remains challenging in particular as the size of the data sets has steadily increased with a wider field-of-view Kelly et al. (2004) and the implementation of laser pulsing capabilities Gault et al. (2006); Cerezo et al. (2007). APT datasets now routinely contain 10s - 100s of millions of ions. Quantifying the microstructural features including their composition, volume fraction, size, geometry, or spatial distribution of secondary phases, requires first some form of segmentation of the APT data Vaumousse et al. (2003). Indeed, chemical segregation means that atoms of certain species are on average closer together, i.e., the local density of atoms of a specific species can be used to distinguish microstructural features.

The most common approach in this direction is the use of iso-concentration surfaces, based on a marching cubes algorithm Lorensen & Cline (1987), that delineate microstructural fea-tures above a threshold composition compared to their surrounding, facilitating visualization. Iso-concentration surfaces provide quantitative information about the composition, and qual-itatively define the geometry and distribution of the microstructural features. The caveat is that, as with most aspects of APT data processing, the information from isosurfaces is very sensitive to user-defined parameters. These parameters are often chosen ad hoc, leading to inconsistencies when the dataset is analyzed by different people Exertier et al. (2018); Barton et al. (2019); Dong et al. (2019).

More recently, various clustering algorithms have been proposed as a more controlled alter-native, for instance, GMM (Gaussian mixture model) Zelenty et al. (2017), DBSCAN (Density-Based Spatial Clustering of Applications with Noise) Stephenson et al. (2007), HDBSCAN (Hierarchical DBSCAN) Ghamarian & Marquis (2019), and OPTICS Wang et al. (2019). Clustering algorithms applied directly to the atomic coordinates of APT datasets tend to be computationally expensive due to the huge number of atoms. Therefore, the impact of the clustering algorithm's hyper-parameters (e.g. number of components or clusters in GMM) on the outcome of this process is rarely evaluated systematically.

Additionally, the morphology of microstructural regions of interest can be geometrically complex, another hurdle for human-guided quantification methods. In recent years, machine learning algorithms have been used more often in the analysis of APT data, with the aim of enhancing automation, reliability, and efficiency of the analysis process. Madireddy et al. presented a deep learning-based edge detection to identify the interface between a matrix and a precipitate phase Madireddy et al. (2019). Although this technique offers a scalable (high-throughput) alternative to iso-concentration surfaces, it does not provide any information about the shape and geometry of the precipitate phase. Peng et al. quantified segregation at grain boundaries through a clustering algorithm applied on spatial coordinates of atoms Peng et al. (2019), followed by calculating composition and thickness fluctuations. Zhou et al. introduced a more precise approach to identify junctions and straight segments in connected interface networks, notably grain boundaries, via deep neural-network image recognition in 2D projections Zhou et al. (2022). The approach works best for columnar grains projected into their basal plane. An extension to interface networks with random orientation, which lack a common projection plane, is unfortunately not straightforward.

Here, we propose a fast approach based on unsupervised machine learning to semi-automatically extract chemical domains corresponding to different phases and segregation zones in the APT data. Contrary to performing clustering on the spatial coordinates of the atoms, we perform clustering in a space containing local composition information of the APT data, with the possibility to also provide the atom clusters in the point cloud within a few minutes on a single CPU core even for datasets containing 500 million atoms. We showcase this approach in the analysis of Fe-doped Sm-Co hard magnets, whose magnetic properties depend on their microstructure. The application of the developed workflows helps to identify and segment different phases into separate precipitates and further disentangle their geometry into plate-like structures enabling meaningful compositional and morphology quantification.

# Materials and Methods

### *Experimental Data*

The production-grade Fe-doped Sm-Co alloy that has been analyzed here is prepared by first milling and crushing the book mold ingots and subsequently mixing the prepared powders to achieve the desired composition (wt%) $Sm_{25}Zr_3Co_{49}Fe_{19}Cu_5$. This is followed by isostatic pressing and sintering of the powders. For further details see Sample synthesis in Duerrschnabel et al. (2017).

Specimens for APT were prepared by focused ion beam on a Dual-Beam Helios Nanolab 600i System using the approach outlined in Thompson et al. (2007). APT measurements were performed on a CAMECA local electrode atom probe (LEAP 5000 XS) in laser pulsing mode, with 10 ps laser pulses at a wavelength of 355 nm (UV), 45 pJ pulse energy at a repetition rate of 200 kHz, in ultra-high vacuum (1 ˆ $10^{-10}$ mbar). Further, the specimen base temperature was kept at 60 K with a detection rate of 6 ´ 10%

### *Workflow*

Figure 1 shows the current overarching workflow to classify different chemical domains in any APT data and to reduce morphologically-complex precipitates into simpler geometries. The spatial coordinates (x,y,z), mass-to-charge state, and ranges defining the elemental nature of each ion are extracted from AP Suite, a proprietary software used to reconstruct and analyze APT data (POS and RRNG files) Day et al. (2019). Further, different chemical domains are extracted from the reconstructed data using clustering in composition space, which is discussed in detail in Section **Classification of chemical domains: clustering in composition space**. After a DBSCAN-based post-processing step, the morphology of the individual chemical domains is considered. If the APT reconstruction contains precipitates with a complex morphology they are disentangled into plate-like simpler structures using a local PCA approach or a U-Net model, which are described in Section **Decomposing geometrically complex precipitates**. The plate-like precipitates or structures can be further analyzed by calculating the in-plane composition and thickness fluctuations. Finally, descriptors corre-sponding to composition, orientation, and thickness for each phase's precipitates are stored systematically using the hdf5 file format The HDF Group (2022). All codes for reading and analyzing APT data are developed in Python, in a Jupyter environment, an open-source web-based tool used for code development and data visualization Kluyver et al. (2016). ParaView, an open-source scientific visualization tool, is used for additional analysis and visualization Ahrens et al. (2005). The code for the demonstrated workflow presented in this paper is accessible at https://github.com/Alaukiksaxena.

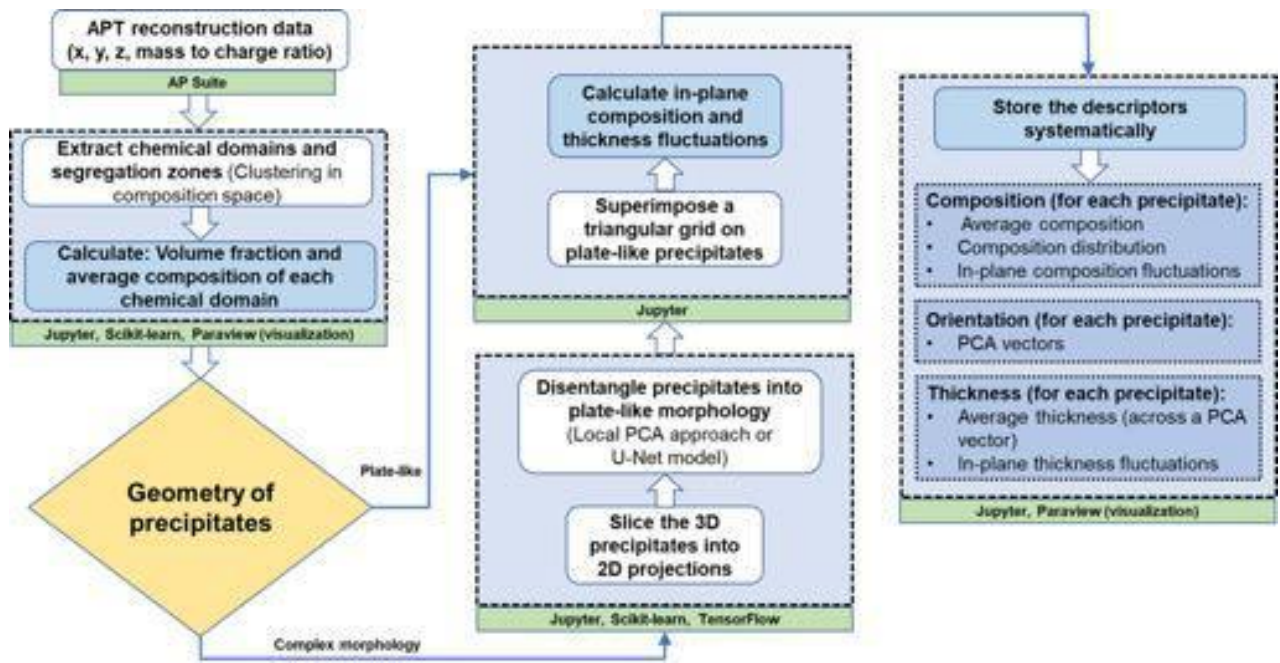

**Fig. 1.** The steps to quantify a 3D microstructure obtained from an APT reconstruction are summarised in the flow chart. The tools and libraries employed in each step are mentioned in the green boxes.

*Classification of chemical domains: clustering in composition space*

First, to identify the chemical domains present in the given APT dataset, we introduce a composition-informed segmentation algorithm to visualize quickly and statistically understand clusters or precipitates and segregation zones in APT data. The algorithm is summarized in Figure 2. The APT data set is divided into equally-sized voxels (here: $2 \times 2 \times 2 nm^3$). For each voxel, the composition, i.e., the relative fraction of all chemical species, is calculated as a vector, cf. Figure 2(B). These vectors describe points in a multi-dimensional unit simplex which we call composition space whose dimensions depend on the number of relevant chemical species in the APT dataset. In this showcase, the APT dataset consists of five elements (Sm, Co, Fe, Zr, Cu), so the composition space is five-dimensional. The rationale behind the proposed method is that the voxels that are present inside a precipitate or cluster, of a particular phase or segregation zone, will have approximately the same composition and therefore appear as clusters in composition space. The size of voxels is a compromise between spatial resolution (smaller voxels can map smaller features) and reducing statistical noise (larger voxels have less scatter in the compositions and better distinguish chemical domains of close compositions). While we use cubes of side 2 nm in our case study, this should be adapted for other datasets, depending on the complexity of the underlying phases and length scale of phase separation.

The second step is to identify different clusters in the composition space. For this, we use the GMM clustering algorithm Hastie et al. (2009), Figure 2(C). The rationale behind choosing

GMM algorithm is that deviation in composition arises from stochastic effects (e.g. due to in-trinsic disorder in the material or the limited detection efficiency) and in limit of large numbers the underlying distribution becomes normal. The GMM algorithm cannot automatically guess the number of clusters in a given dataset. This information has to be fed to the algorithm as a hyperparameter. The approximate number of clusters or chemical domains in the composition space can be determined using the Bayesian information criterion (BIC) or can be inferred from, e.g., thermodynamic databases. In the current work, we have used minimization of the BIC to get the number of clusters as shown in Figure 3(A). For the given APT datasets, we get a visible kink or elbow, i.e., the number of clusters after which the decrease in the BIC value is not significant (threshold derivative of BIC = ˊ0.1 ˆ $10^6$), at three clusters, which physically correspond to the 2:17 phase (matrix), Z phase (rich in Zr) and 1:5 phase (rich in Cu) Zelenty et al. (2017) Geron (2019). The number of clusters is also in agreement with the known phases of this material Duerrschnabel et al. (2017). The kink can be more clearly seen in the central difference numerical derivative of the BIC as shown in Figure 3(B). Figure 4(B), shows a 3-D (Cu, Fe, Sm) projection of the 5-D composition space where each point corresponds to a voxel in the spatial coordinate system (refer to Supplementary video 1). The three clusters or chem-ical domains found by the GMM algorithm are marked in Figure 4(B). Further, normalized frequencies of voxel composition for each chemical species shown in Figures 4(A, C, D, E, F), obtained using kernel density estimate, highlight that no single element is sufficient to assign a voxel to a certain phase, but definitely their combination is required. The voxels pertaining to a particular cluster in composition space are translated back to the spatial coordinate system so that atoms present in the matrix, precipitates, and segregation zones can be separated, Figure 2(D).

Further, DBSCAN is applied to the centroids of the voxels present in a particular phase or segregation zone Ester et al. (1996). Here, a voxel centroid is the arithmetic mean of the positions of the atoms present in the given voxel. In this process, the noisy voxels, which often lie at the boundary of the dataset, are removed and various precipitates pertaining to each phase are identified as shown in Figures 2(D and E) (refer to Supplementary video 2). Since both the atoms residing in each voxel and the voxels present in each precipitate are registered, the last step enables us to study each precipitate separately. Moving from an atom-by-atom representation of the data set to voxel centers is particularly advantageous for DBSCAN since the atomic density fluctuations due to the APT experiment (e.g., pole formation Gault et al. (2012b) or in-homogeneous evaporation due to evaporation-field differences Vurpillot et al. (2000)) can be smoothed out. The voxel centers, by construction, are equally spaced and provide a practically constant density of points within each precipitate. Since now the number of voxels that are present in each precipitate is known, statistical descriptors like mean volume, number density, and nearest neighbor distribution of the precipitates can be readily obtained.

## *Decomposing geometrically complex precipitates*

**Local principal component analysis(PCA)-based method**: In APT datasets, especially while studying nano-grain structures, we often encounter quasi-planar interfaces or interphase boundaries that cross each other forming triple or quadruple junctions. This geometric com-plexity makes the quantification of such microstructures cumbersome and challenging. In the Sm-Co alloy under our investigation, the Cu-rich 1:5 phase has precipitates that do not cross the entire field of view, making the approach of Zhou et al. Zhou et al. (2022) discussed in Section **Introduction** impossible to use. To separate complex precipitates into plate-like substructures visible to the human eye, we propose a local PCA-based method. This method is applied to all identified precipitates to find plate-like structures automatically.

This method is demonstrated on a 1:5 phase precipitate, as shown in Figure 5(A), with an entangled morphology. First, PCA is performed on the spatial coordinates of voxel centroids belonging to the chosen precipitate, and all the voxel centroids are transformed to the PCA coordinate system (PC1, PC2, PC3), Figure 5(B). For visualisation a 2D projection of the 3D precipitate in the PC3 direction is shown. Next, the precipitate is divided into $N_{cut}$ slices, shown by the dotted lines in the figure, along the PC1 direction. $N_{cut}$ depends on the thickness of a precipitate. Each slice is projected in the PC1 direction, i.e., to the PC2-PC3 plane, Figure 5(B). The slicing process is repeated in the PC2 and PC3 directions to generate 2D projections in the respective perpendicular planes. The created 2D projections are shown in Figure 5(C) and will be used to identify the junctions connecting the ribbon-like features in the projection plane.

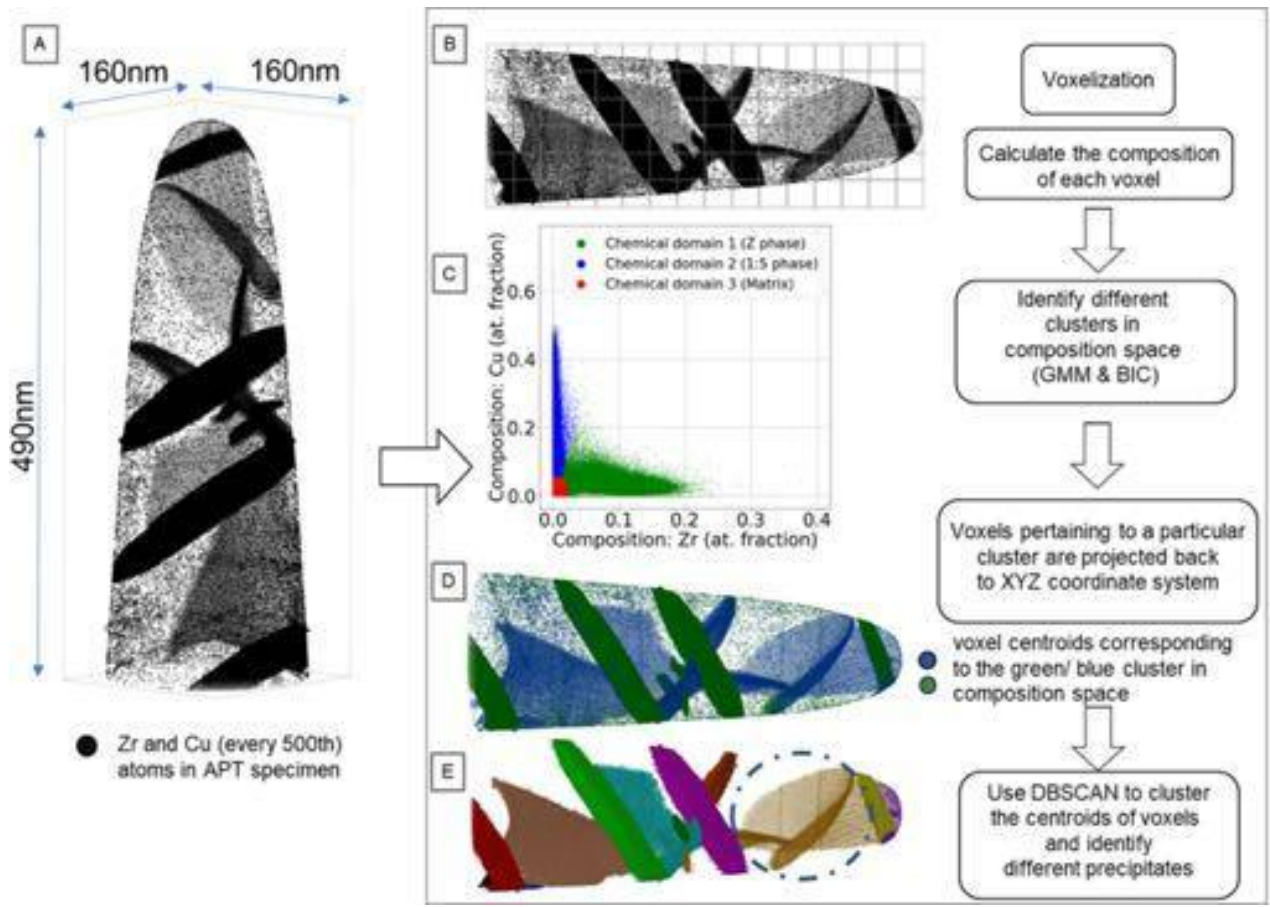

**Fig. 2. (A)** APT reconstruction for a Fe-doped Sm-Co alloy. For clarity, only minority elements (Cu, Zr) are shown. **(B)** Voxelization and per-voxel composition analysis, the grey grid schematically illustrate the voxels. **(C)** 2D projection of the 5D composition space. Each point corresponds to one voxel. Color coding according to 3 composition clusters identified (Chemical domain 1 (green), Chemical domain 2 (blue), and Chemical domain 3 (red) ). **(D)** Phases in real space at voxel resolution according to composition classification. **(E)** DBSCAN is used to cluster voxel centroids belonging to each phase to recognise separate individual precipitates. The circled precipitate has a complex morphology.

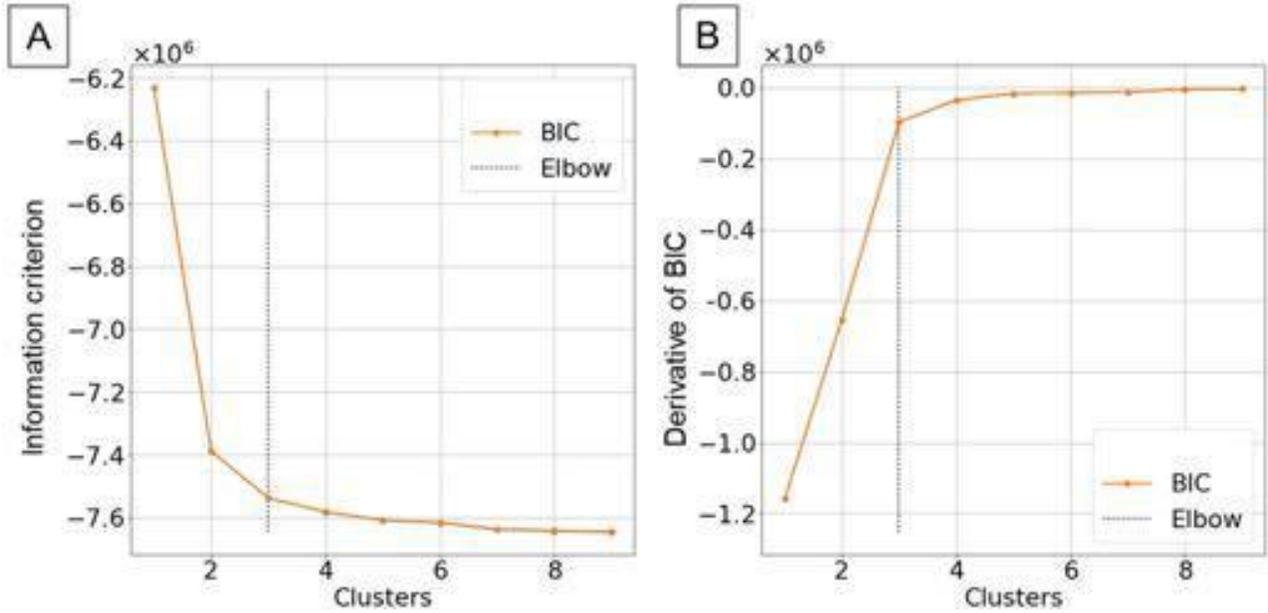

**Fig. 3. (A)** An elbow, shown with a dotted line, at clusters = 3 is observed in the BIC curve. The number of clusters is in agreement with the known phases of the given material. **(B)** The central difference numerical derivative of BIC is invariant from clusters = 4.

For this, the local PCA approach is applied to each of the 2D projections or slices. This method helps in the classification of voxel centroids (data points in the 2D projection) to be in junction-like or non-junction-like environments, Figure 5(D) to (F). The local PCA algorithm is summarized in Figures 6(A) to (D). For each data point in a 2D projection (e.g., the orange one), we find all neighborhood points lying within a radius $r$ (green) and, from the PCA of the distance vectors, identify the direction of maximum variance, $PC1_{center}$, Figures 6(A) and (B). The radius $r$ is provided as a hyperparameter for each precipitate. The $PC1_{center}$ direction is then compared to the corresponding one, $PC1_{neigh}$, for all the data points in the neighborhood (green) by computing the dot product ($P\,C1_{center} \cdot P\,C1_{neigh}i$ where i is a given neighborhood data point). If the selected data point (orange) lies in a quasi-plate-like structure then most of the $P\,C1_{neigh}$ directions will point in the same direction as $P\,C1_{center}$ and the average of their dot products will be close to one. However, if the orange data point lies near the junction then most of the $P\,C1_{neigh}$ directions will not point along $P\,C1_{center}$ and the average of their dot product is lower (generally, 0.4 to 0.8 depending upon the geometry), Figure 6(C). In this way, the local PCA approach, which is applied to all the data points in a 2-D slice, is able to classify the data points that lie at the junction of the precipitates, Figure 6(D). All the junction data points from each 2-D slice are stored in a list and tracked back to the original points in the 3-D spatial coordinate system. Finally, the junction centroids in 3D space are removed from the set of all the voxel centroids pertaining to a particular precipitate, separating the quasi-plate-like structures. These structures can now be identified as separate clusters by applying DBSCAN since the connecting junction between them has been removed, Figure 5(F).

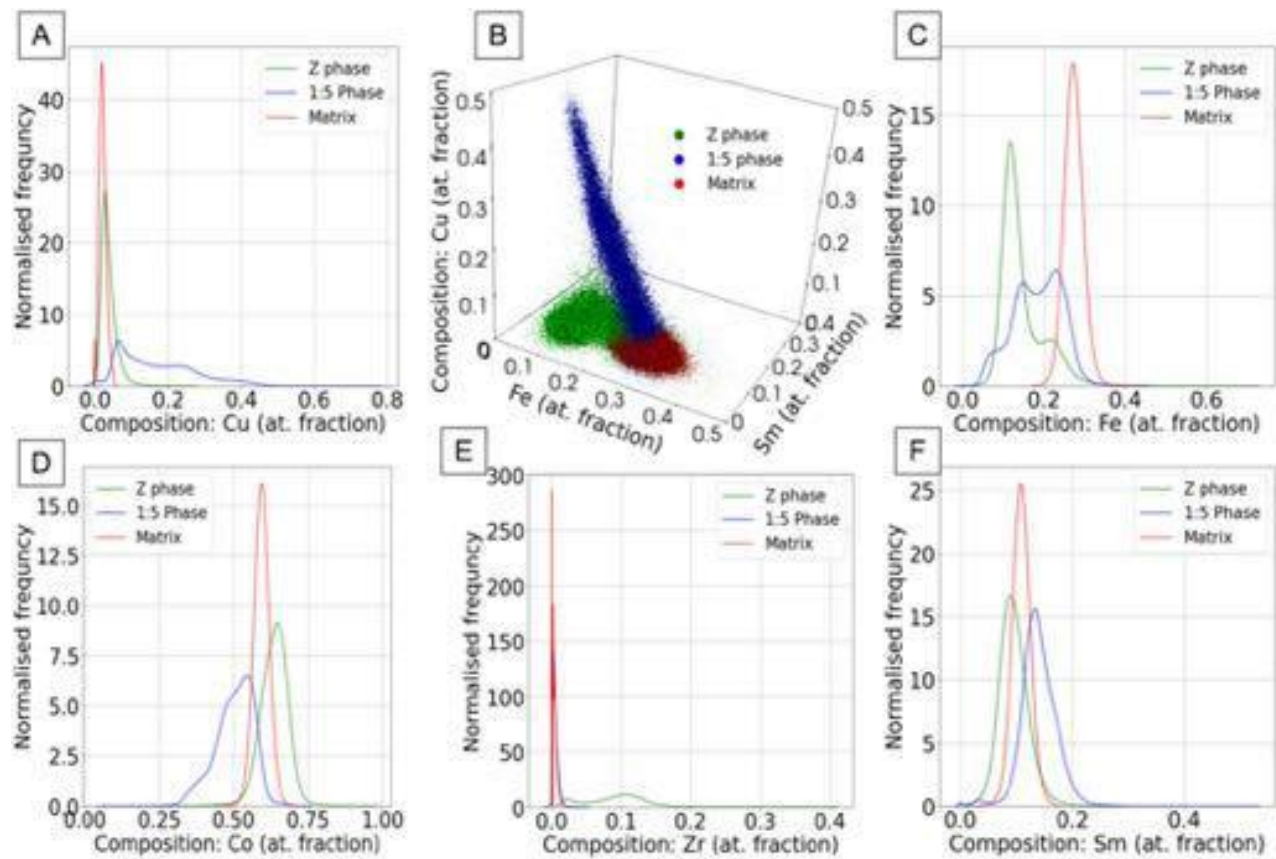

**Fig. 4.** Normalized frequencies of voxel composition (at. fraction) of **(A)** Cu, **(C)** Fe, **(D)** Co, **(E)** Zr, and **(F)** Sm, respectively, for each phase present in one of the Sm-Co alloy APT dataset. **(B)** 3D projection (Cu-Fe-Sm composition) of the 5D composition space with clusters pertaining to respective phases.

**U-Net based approach**: In the local PCA-based approach, the accuracy of the classification of data points as junction depends on the hyperparameter *r*, which has to be fed for each precipitate, shown in Figure 6(A). For example, if the radius *r* is larger than the precipitate the local PCA approach will not be able to find the data points at the junction. To circumvent this problem, we have developed a more robust approach based on supervised machine learning (U-Net) and image augmentation. A U-Net is a convolutional neural network (CNN) based architecture, which is used in computer vision problems, especially biomedical image data, for semantic segmentation Ronneberger et al. (2015). Semantic segmentation means that the deep learning algorithm predicts the class for each pixel in a given image.

First, the local PCA approach is applied to 1:5 phase precipitates extracted from a given APT dataset. This process created 500 2D projections or slices. The data points in the projections are already labeled as junction or not junction using the local PCA approach. All the projections are converted to gray-scale images of shape $512 \times 512 \times 1$ and based on the label information a mask depicting the junction in each image is also extracted with shape 512 ˆ 512 ˆ 2, where the last 2 channels correspond to the classes: junction and not junction.

To increase the diversity of the training data, image augmentation steps like rotation, translation, flipping and resizing are applied to the original images to create a pool of 3000 training images and their corresponding masks. These training images (512 ˆ 512 ˆ 1) and the masks (512 ˆ 512 ˆ 2) are used to train a U-Net. The trained model, as shown in Figures 6(E) to (G), takes an image corresponding to a projection as input (512 ˆ 512 ˆ 1) and predicts a mask of shape 512 ˆ 512 ˆ 2, where the last channel contains the probability of a pixel to be a junction. Based on the predicted mask, the data points in the original projection/slices are classified as junction or not.

The U-Net based model is more robust and is directly applied to the other samples of the same material without the need to tune the parameter *r* for each precipitate. The local PCA-based approach is semi-automatic and can be tuned according to the precipitate size. However, the U-Net-based approach is fully automatic irrespective of the precipitate size. Nevertheless, retraining of the U-Net model might be required when applying the same model to other material datasets, in which the geometry of precipitates is different. Further details of the U-Net architecture used in this study are given in the supplementary Figure 12.

## Results and Discussion

### *Calculation of volume fraction and composition of each phase:*

The segmentation of each phase or chemical domain allows for the calculation of the approx-imate volume fraction, as the number of voxels included in a phase is known. The volume fraction of each phase in one of the APT datasets is summarised in Table 1. The average composition of a phase is the mean of the compositions of the precipitates or domains it com-prises. Average compositions and standard deviation for each phase in the APT sample are summarised in, Table 1.

**Composition fluctuations in precipitates**: The comparison between the voxel compo-

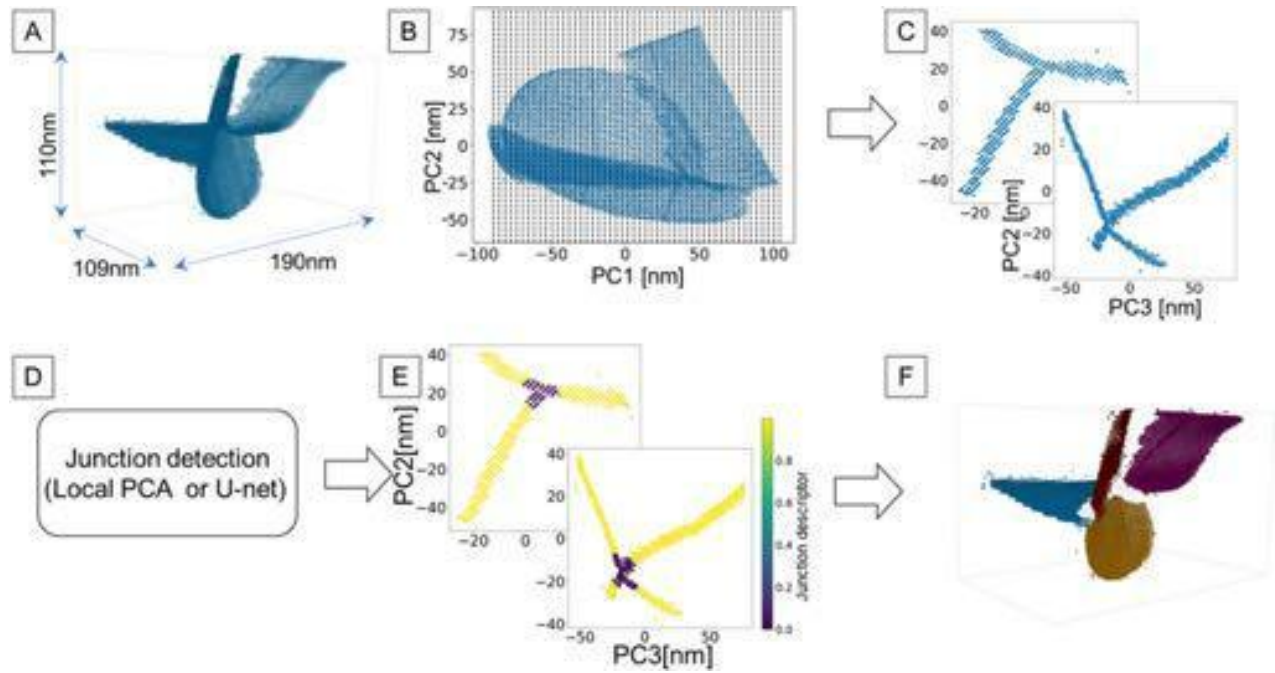

**Fig. 5.**    Workflow for precipitate disentanglement via junction detection and removal.

**(A)**  Voxel centroids corresponding to a 1:5 phase precipitate having four plate-like structures meeting at a junction, also shown in Figure 2(E). **(B)** The voxel centroids are transformed to the PCA coordinate system and divided into $N_{cut}$ slices along PC1 (direction of maximum data variance). For visualisation a 2D projection of the 3D precipitate in PC3 direction is shown. **(C)** 2D projections for slices of voxel centroids in the plane perpendicular to the cutting direction. **(D)** and **(E)** Local PCA approach or U-Net model trained on similar data is used to detect junctions in the 2D projections. **(F)** All the voxel centroids that are labeled as plate-like back-projected to 3D space where they are clustered using DBSCAN.

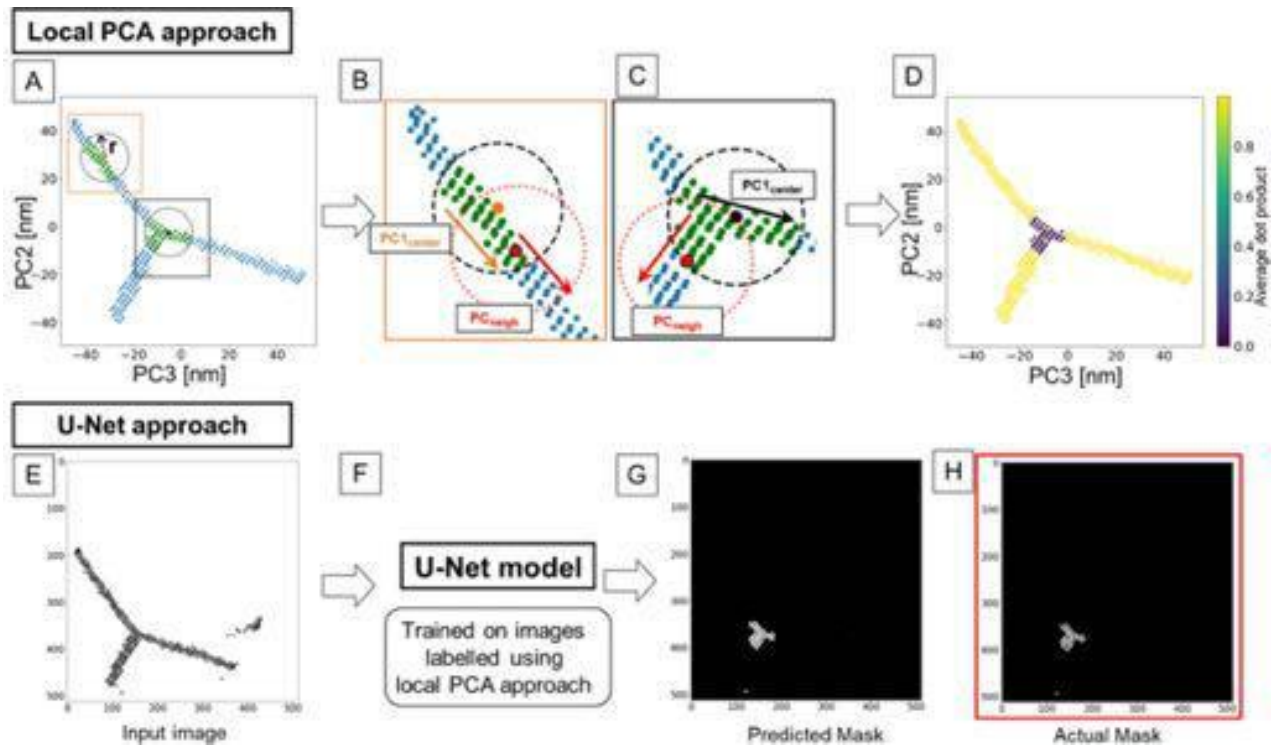

**Fig. 6. Junction detection. (A)** A 2D projection of the voxel centroids in a slice extracted from a 3D precipitate. The orange and the black box highlight a plate-like and junction region, respectively. The neighborhood (green) within a radius $r$ around a chosen point (orange) is selected within the orange box. **(B)** Magnified region away from the junction. The direction of maximum variance, $PC1_{center}$ and $PC1_{neigh}$ corresponding to the neighborhood of the orange and red points within radius $r$, respectively. The red point is chosen from the neighborhood of the orange point. **(C)** Magnified region near the junction. **(D)** Average dot product mapped on the 2D projection delineating the junction region. **(E) to (H)** A U-Net model trained on local PCA approach labeled images is applied on a previously unseen input image to predict the junction region or mask. The actual or true mask is also shown.

**Table 1.** Volume fraction, average thickness and average composition of 2:17, Z, 1:5 phase in one of the given APT datasets.

| Phase | Volume fraction | Thickness(nm) | Composition at% | | | | |
|-------|-----------------|---------------|------|------|------|------|------|
| | | | Sm | Co | Fe | Cu | Zr |
| 2:17 | 0.8 | - | 10.9 ˇ 0.2 | 59.4 ˇ 0.5 | 27.2 ˇ 0.4 | 2.0 ˇ 0.4 | 0.5 ˇ 0.1 |
| 1:5 | 0.06 | 17.6 ˇ 4.0 | 14.4 ˇ 0.8 | 48.0 ˇ 2.7 | 16.1 ˇ 2.6 | 21.1 ˇ 4.5 | 0.4 ˇ 0.1 |
| Z | 0.14 | 10.9 ˇ 1.4 | 9.7 ˇ 0.7 | 63.4 ˇ 1.3 | 14.5 ˇ 0.7 | 3.8 ˇ 0.7 | 8.5 ˇ 0.9 |

sition distribution of each chemical species in a particular precipitate and the corresponding random distribution are plotted to understand if any clusters or segregation zones are present inside the chosen precipitate. Such a comparison for randomly selected precipitates of the 1:5 and Z phase and a cell of 2:17 matrix phase for Cu, Fe, and Zr is shown in Figure 7. As seen in the figure, the experimental distribution deviates significantly from the random distribution for the Z and 1:5 phases as compared to the 2:17 phase. This behavior is seen across all APT data sets of the current material. In order to understand and quantify these deviations, all the precipitates are reduced to simpler plate-like structures enabling us to study in-plane composition fluctuations.

### *Comparison with isosurface results:*

The proposed approach of clustering in composition space has been applied to all the given Fe-doped Sm-Co alloy APT datasets. Figure 8 provides a visual comparison between the isosurfaces obtained using AP suite and the results of our new algorithm. Figure 8(A), shows the 6 at% Zr (Z phase) and 12 at% Cu (1:5 phase) isosurfaces, and Figure 8(B) shows the voxel centroids that lie in Z phase and 1:5 phase. These phases have been identified by the composition space clustering, while the final assignment of the voxels as part of an extended precipitate additionally relies on the DBSCAN algorithm on the voxel centroids of each phase to eliminate noise. The phases obtained in Figures 8(B) are visually identical to the isosurfaces for the respective phases shown in Figure 8(A)

For clearer visualization of the precipitates, we separated them in subfigures (C) and (D): Figure 8(C) depicts only the 1:5 phase precipitates, and Figure 8(D) the ones for the Z phase. The precipitates of 1:5 phase with complex morphologies containing bi and quad junctions are highlighted in Figure 8(B). The junction voxel-centroids have been identified using the U-Net approach and removed (Section **Decomposing geometrically complex precipitates**). The remaining voxel centroids are clustered employing DBSCAN to obtain plate-like precipitates as shown in Figure 8(C) (refer to Supplementary Videos 2 and 3).

### *In-plane composition and thickness fluctuations*

We modified the algorithm of Peng et al. to find the in-plane composition and thickness fluctuations in the plate-like precipitates Peng et al. (2019). The first step is to apply a PCA transformation on the voxel centroids of a given quasi-planar precipitate. Next, a regular 2-D triangular grid of size 4 nm is superimposed on the precipitate in the plane formed by the two maximum variance PCA directions. The grid triangles not containing voxel centroids are removed. This is followed by moving each node of the trimmed grid to the center of mass of the voxel centroids located in the cuboidal region overlayed at that node, shown in the supplementary Figure 13 (A). We use voxel centroids of the precipitates for analysis while the original algorithm directly uses atoms at the interface to be analyzed. Since some of the precipitates in our case studies have a million atoms or more, the choice of voxel centroids makes it faster. As a result of moving the nodes, the trimmed grid takes the shape of the precipitate, supplementary Figure 13 (B) and (C). The geometry of the precipitate is still preserved if we use voxel centroids for our analysis.

After obtaining the grid that conforms with the shape of the precipitate, the local surface orientation (normal vector) at each node is defined from the average of the adjacent triangles. This is followed by selecting atoms in a cylindrical region of interest (ROI) at each node with the axis of the cylinder along the node's normal. To speed up the atom search, we exploit the space partitioning provided by the voxels. First, relevant precipitate voxels are identified, whose centroids falls inside a cylindrical ROI with a radius, $r_{out}$ „ 4 nmq, and length $L_{out}$ „ 34 nmq. From the selected voxels, the atoms within a cylindrical ROI of dimensions, $r_{in}$ „ 2 nmq, and length $L_{in}$ „ 30 nmq, are chosen for the calculation of in-plane thickness and composition fluctuations.

Each cylindrical ROI is divided into bins along the cylinder axis and the composition of each bin is calculated. We arbitrarily selected a bin width of $L_{bin}$ „ 0.6 $nm$q as it provided a good compromise of limited statistical fluctuation and details of the feature of interest. The 1D composition profile for all the elements, obtained at a certain node in a precipitate of the Z phase along the normal shown in the Figure 9 (A), is plotted in Figure 9 (B). To analyze the clustering or segregation of the chemical species across the quasi-planar precipitates, we study the 1D composition profiles in more detail. To exemplify our approach, Figure 9(C) presents a 1D composition profile of Cu across the aforementioned node in the Z-phase precipitate. A six-degree polynomial is fitted on the 1D composition profile. Subsequently, the number and location of local maxima in the fitted polynomial are obtained using find_peaks, a tool implemented in open source scipy library Virtanen et al. (2020). Figure 9(A) shows the fluctuations in the number of modes (or peaks) in 1D Cu composition profiles of the Z phase precipitates in one of the APT data sets. Almost all the precipitates have 2 or 3 modes in their composition profiles which suggests that either Cu is segregated at the edges or it is also segregated at the middle of the precipitates.

Based on the 1D composition profiles at each node, it is possible to reveal possible in-plane composition fluctuations, as shown in Figure 10 for Zr (A and B) and Cu (C and D) for precipitates of Z (A and C) and 1:5 phase(B and D). Since the composition of Co is significantly higher than any other element in both Z and 1:5 phases (refer Table 1) and also its composition is higher in the Z phase and lower in the 1:5 phase compared to the matrix (2:17 phase), it is the best choice to mark the start and end of a precipitate. The in-plane fluctuations are calculated by taking the average of an element's composition over the length of the ROI where Co composition is not zero. As shown in Figure 9 (B), the average is calculated over the bins lying between $Dist_{max}$pCoq and $Dist_{min}$ pCoq. This line of reasoning is also used to calculate the thickness of each precipitate at every node of the grid. Thickness, T = $Dist_{max}$pCoq ´ $Dist_{min}$pCoq, of the Z precipitate at the node, is shown in Figure 9(B). Thickness fluctuations for precipitates of Z and 1:5 phase are summarised in supplementary Figure 14(B) and (C).

### *Quantification of the microstructure*

To quantify the Sm-Co alloy microstructure, in addition to the thickness, orientation, and spacing between the extracted precipitates have to be calculated. The parameters like the distance between parallel precipitates and the angle between the intersecting ones are used to define a 2D model microstructure of a Sm-Co alloy Katter et al. (1996). Similarly, we have extended this concept to quantify 3D microstructures. To define the orientation of each plate-like precipitate, PCA is carried out for the spatial coordinates of voxel-centroids present in the given precipitate. The direction of minimum variance in the spatial coordinates is then taken as an approximate orientation normal to the precipitate plane. Figure 11(A) shows an example. The 1:5 phase precipitates with complex morphology in the given microstructure have been reduced to plate-like structures. The angle between each pair of planar precipitates is calculated and a set of parallel precipitates (the angle between the normal vectors is approx-imately 0) forming the magnetic domain walls is shown in Figure 11(B). Similarly, a set of parallel Z phase precipitates in the same APT sample is shown in Figure 11(C). The distance between a pair of parallel precipitates is the distance of the centroid of one of the precipitates from the plane of the other precipitate. The distances between the consecutive Z phase plates, and a set of parallel 1:5 phase precipitates are summarised in Figure 11(B) and (C) (refer to Supplementary Video 4).

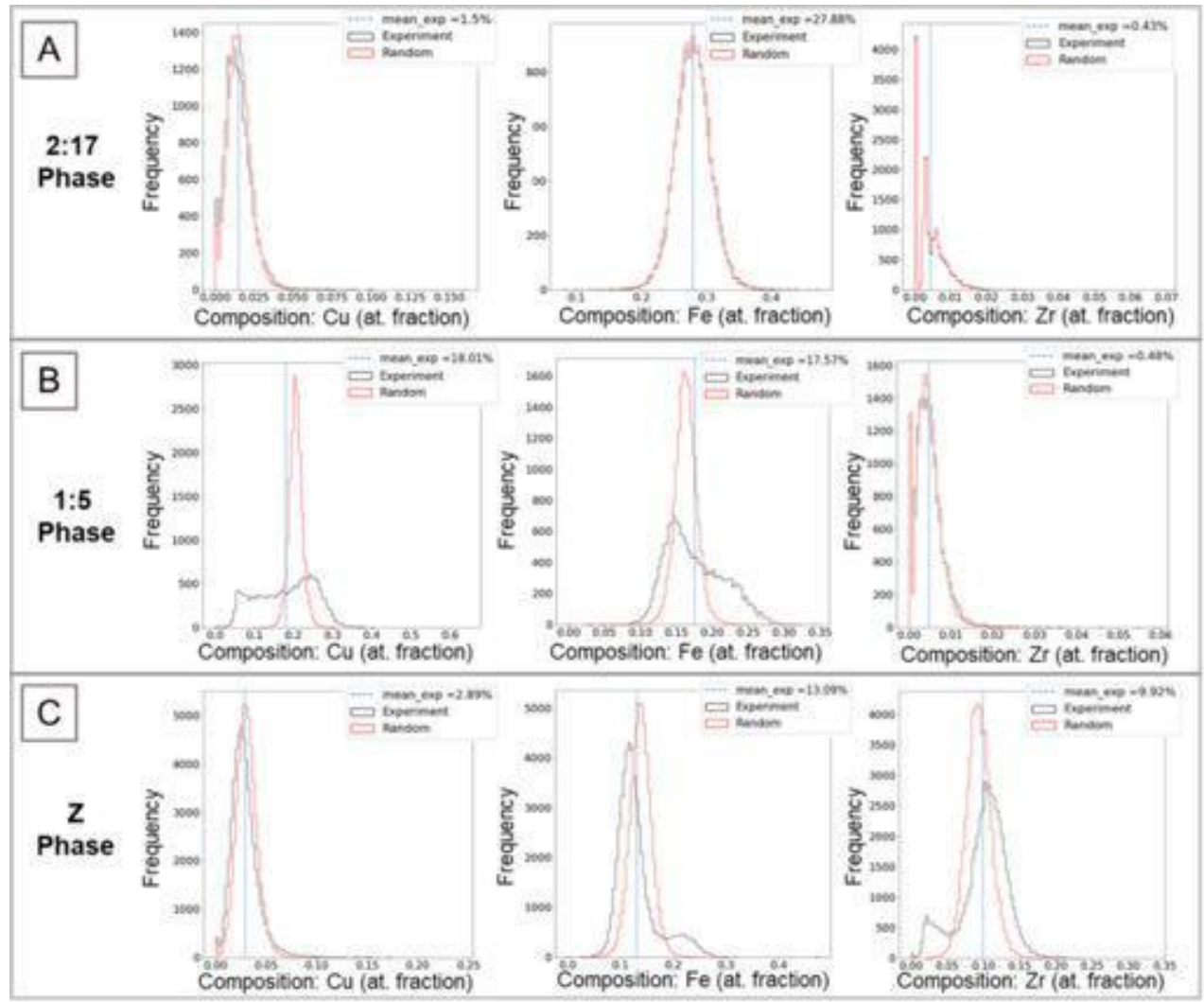

**Fig. 7.** Voxel composition distributions of Cu, Fe and Zr present in **(A)** a 2:17 matrix cell as well as in 2 precipitates chosen for **(B)** 1:5 phase and **(C)** Z phase. The experimental composition distribution (black) deviates significantly from the random distribution (red) for 1:5 and Z precipitates.

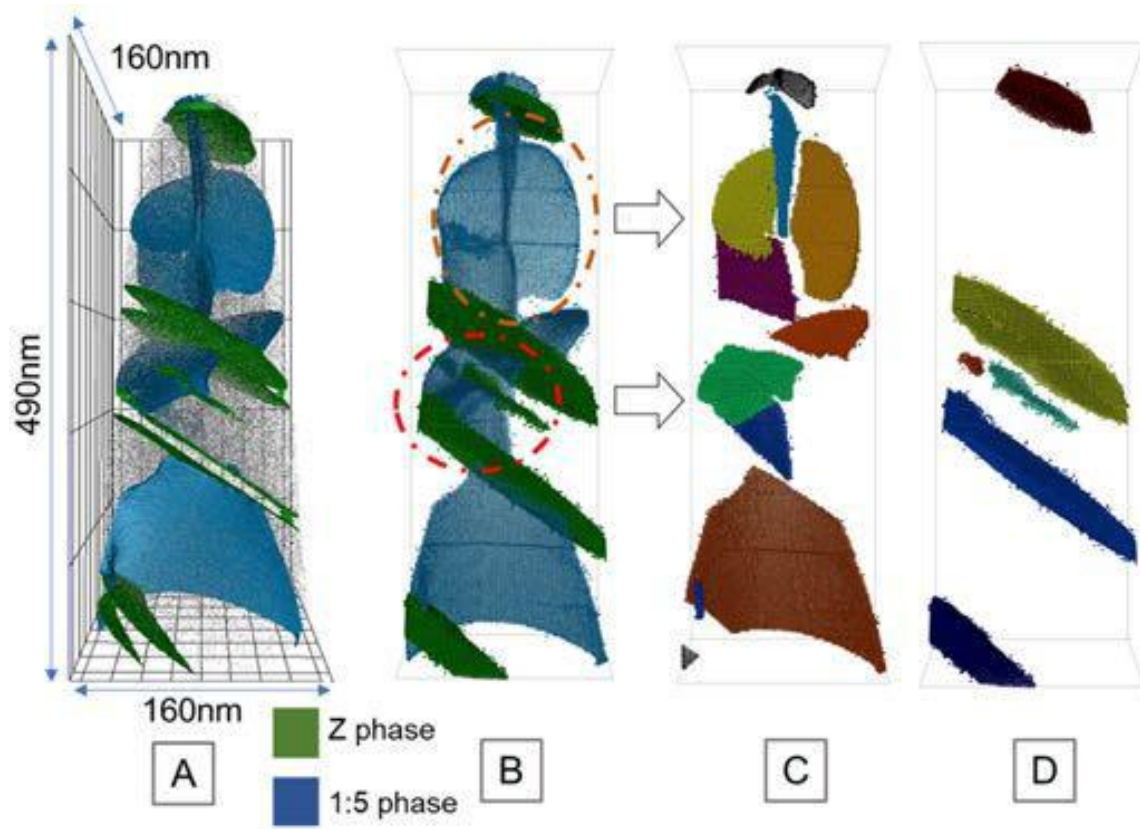

**Fig. 8. (A)** 6 at% Zr (green: Z phase) and 12 at% Cu (blue: 1:5 phase) isosurfaces in a Fe-doped Sm-Co alloy APT dataset extracted using AP suite. **(B)** The voxel centroids lying in the Z phase (green) and 1:5 phase (blue) precipitates extracted using clustering in composition space. The 1:5 phase precipitates with complex morphology are highlighted. **(C)** Segmented 1:5 precipitates. Complex morphology precipitates are separated into plate-like structures. **(D)** Segmented precipitates of the Z phase.

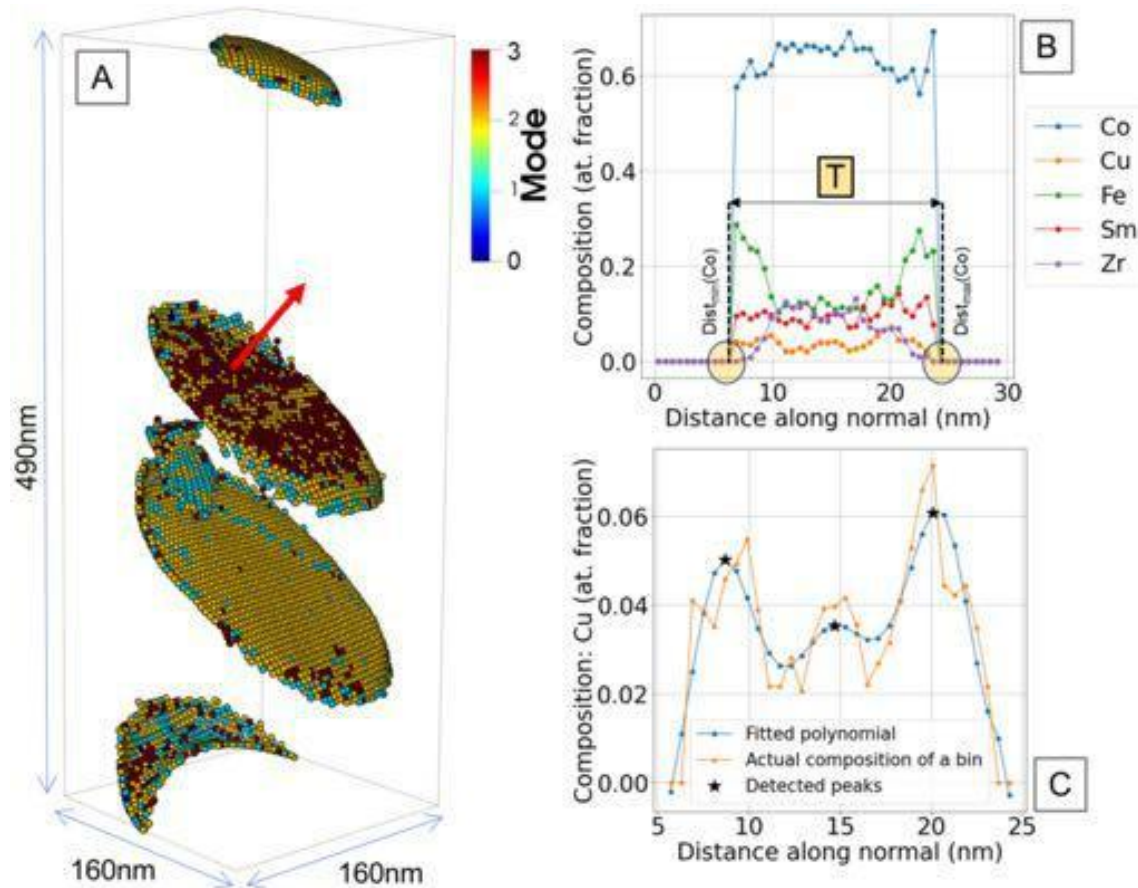

**Fig. 9. (A)** The nodes in a triangular grid superimposed on the Z phase precipitates. The color of each node pertains to the number of modes or peaks in the 1D composition profile of Cu at that particular node. **(B)** 1D composition profile of Co, Cu, Fe, Sm, and Zr along a cylindrical ROI around the red arrow(normal at a node) shown in panel **(A)**. The thickness (T) of the precipitate along the ROI is estimated by calculating the maximum distance between the bins that have non-zero Co composition.**(C)** 1D composition (at. fraction) profile (orange) of Cu along the same ROI. A six-degree polynomial (blue) is fitted on the 1D composition profile. Local maxima or peaks (black) are estimated in the fitted polynomial to approximate the local segregation zones of Cu across the precipitate.

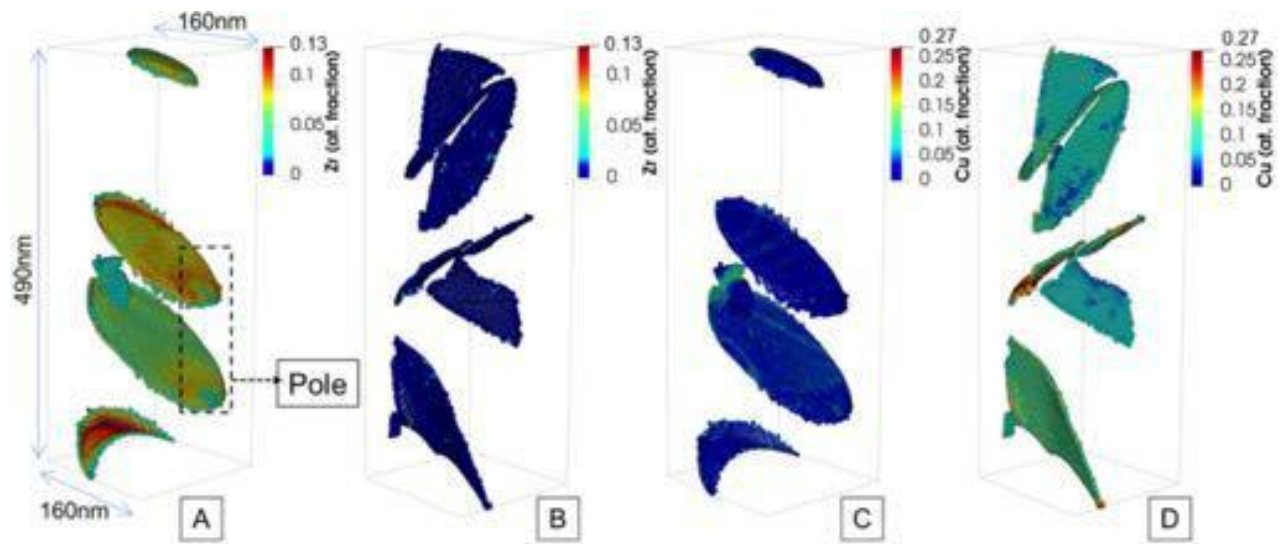

**Fig. 10. In-plane composition fluctuations**: The nodes associated with 2D grids fitted on plate-like precipitates in one of the APT samples. The color coding corresponds to the average composition (at. fraction) of Zr (**A**, **B**) and Cu (**C**, **D**) at each node in Z phase(**A**, **C**) and 1:5 phase (**B**, **D**). Further, in panel **(A)**, Zr is depleted in a circular region in two precipitates, marked with a rectangular box. This can be a pole in the dataset, which is an APT artifact.

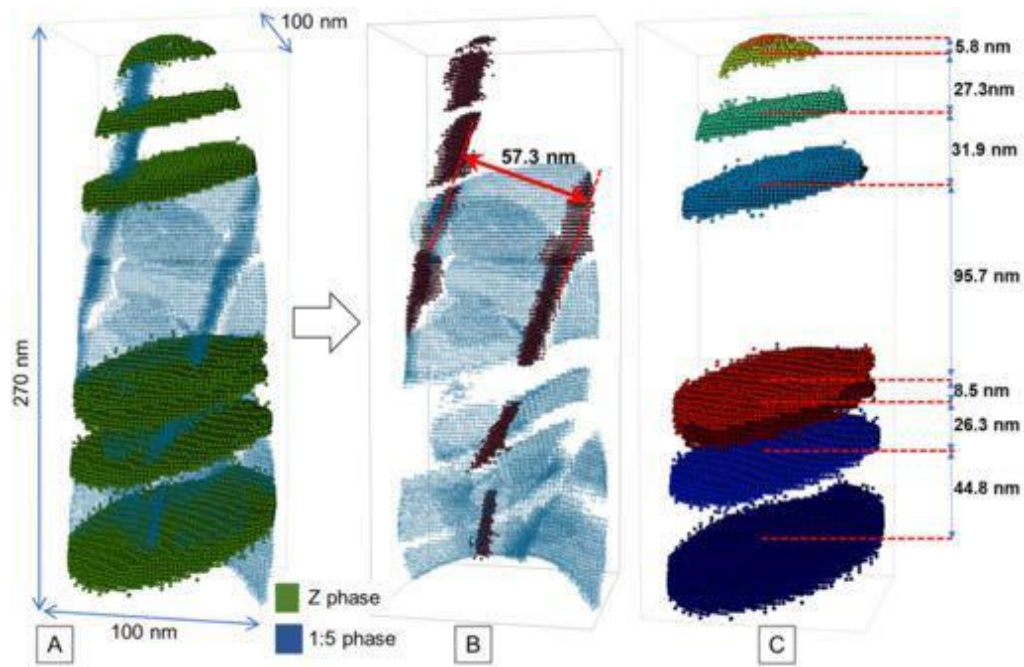

**Fig. 11. (A)** Voxel-centroids of Z and 1:5 phase, green and blue respectively, present in one of the APT samples with complex microstructure. **(B)** and **(C)**Sets of parallel 1:5 and Z phase precipitates, respectively, with distances between them.

# Summary and Conclusion

In this paper, we developed a python based ML workflow to chemically and geometrically quantify the 3D microstructure from APT data. The first part of the workflow extracts existing chemical domains or segregation zones. The key steps are (1) voxelisation of the APT data set, (2) calculating the composition of each voxel, (3) clustering in the composition space to obtain the chemical domains in the dataset and (4) separation of voxels of the extended regions of each chemical domain by DBSCAN clustering algorithm. The relevant data of all steps is stored in hdf5 files to aid subsequent postprocessing. This enables a robust and systematic analysis of the microstructure from its chemical fingerprint, yielding for instance the average volume fraction, domain size distribution, and average composition of the identified chemical domains (which could be stable phases from the thermochemical phase diagram, or segregation zones with a unique composition near grain boundaries or other crystallographic defects).

Geometrically quantifying the microstructure requires to reduce morphologically entangled phases to individual structures. This was achieved here for a particular type, namely agglomerates exhibiting plate-like substructures. Our novel local PCA and/or U-Net approaches allow us to distinguish junction voxels from those lying in the planar regions, followed by a DBSCAN clustering to get the separated planar structures. This procedure can efficiently and rigorously segment out predominantly flat subdomains with arbitrary orientations in 3D. Indeed, planar or plate-like morphologies are among the most common microstructural features, for example, segregation at the interface (grain boundary) and inter-phase boundaries or phases formed in some Al Khushaim et al. (2015); Zandbergen et al. (2015) and Ni alloys Vogel et al. (2015). Then descriptors such as PCA vectors, their explained variance, average composition, cen-troid, or the angle of the inclination with the axis of the APT tip, are calculated for each planar substructure. Using the geometrical information, the compositional analysis can be further refined to yield in-plane composition and thickness fluctuations. A method to map the segregation of solute atoms across interfaces is also developed.

Using the developed workflow, we identified three phases, matrix (2:17 phase), Z phase and 1:5 phase, for Fe-doped Sm-Co alloy APT datasets. Notably, this computational analysis can be executed in just a few minutes using a single CPU core, even for APT datasets with 500 million atoms. We used the workflow to disentangle each precipitate, allowing a fast and robust analysis. We were able to identify the deviation of the experimental distribution of Zr, Fe and Cu in precipitates of Z and 1:5 phase from their corresponding random distributions. The across-plane composition fluctuation of Cu was very distinct with enrichment at the edges or also at the center of the Z-phase precipitates. Such compositional fluctuations together with geometrical quantification of the microstructure are very relevant for the magnetic performance of this and other alloys.

The developed workflow can readily be applied to a wide variety of microstructures analysed by APT. This enables a consistent quantification of chemical and structural features in an APT dataset.

## code Availability

The workflow is available at https://github.com/Alaukiksaxena/CompositionSpaceNFDI

### Acknowledgements

A.S. appreciates funding by Helmholtz School for Data Science in Life, Earth and Energy (HDS-LEE). NP is grateful for funding from the IMPRS SURMAT. SK, NP and BG acknowledge support from the Deutsche Forschungsgemeinschaft (DFG) for funding from the Leibniz Prize 2020 (GA 2450/2-1). OG, LML and BG are grateful to DFG for funding of the TRR 270 HoMMage (INST 163/578-1). NK, SK, BG, JN and CF are grateful for financial support from BiGmax, the Max Planck Society's Research Network on Big-Data-Driven Materials Science.

# Supplementary material

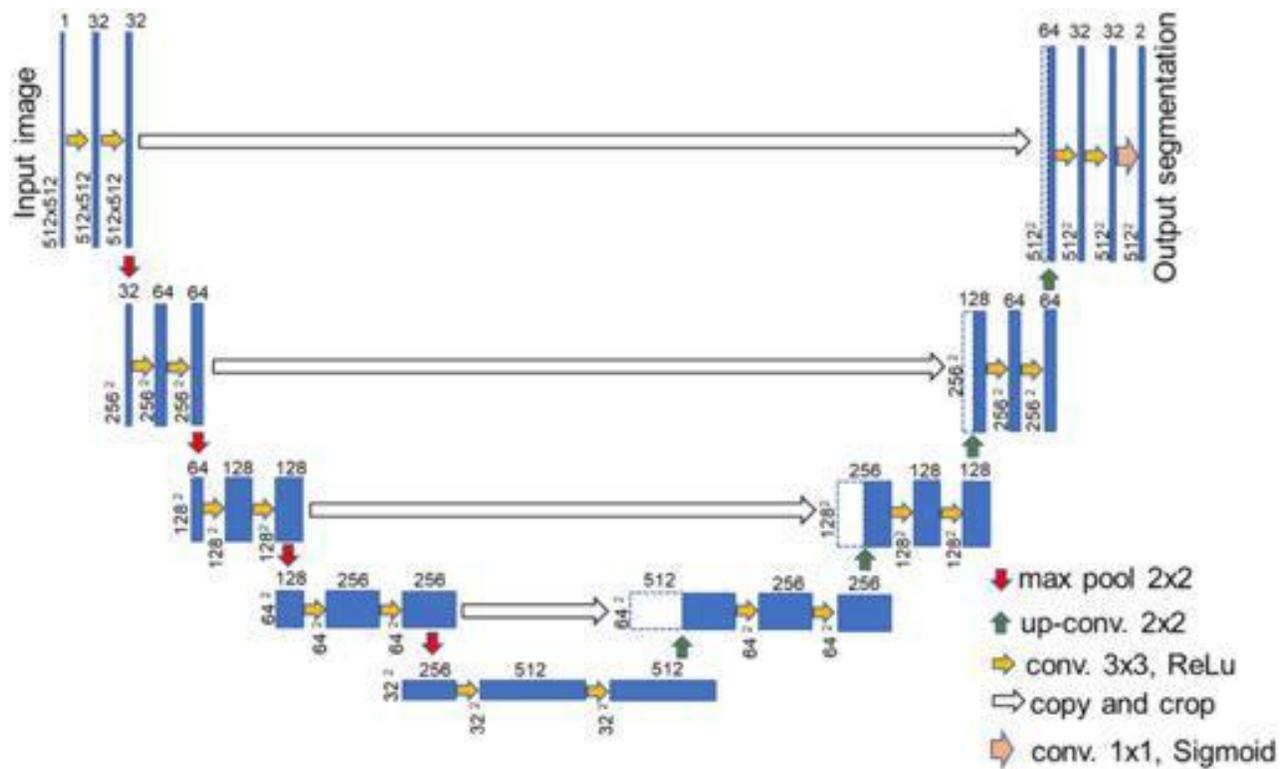

**Fig. 12.** The architecture of U-Net image segmentation model used in this study. Categorical cross-entropy is used as the cost function for the model which is optimized using Adam optimizer (learning rate = $e^{-5}$)

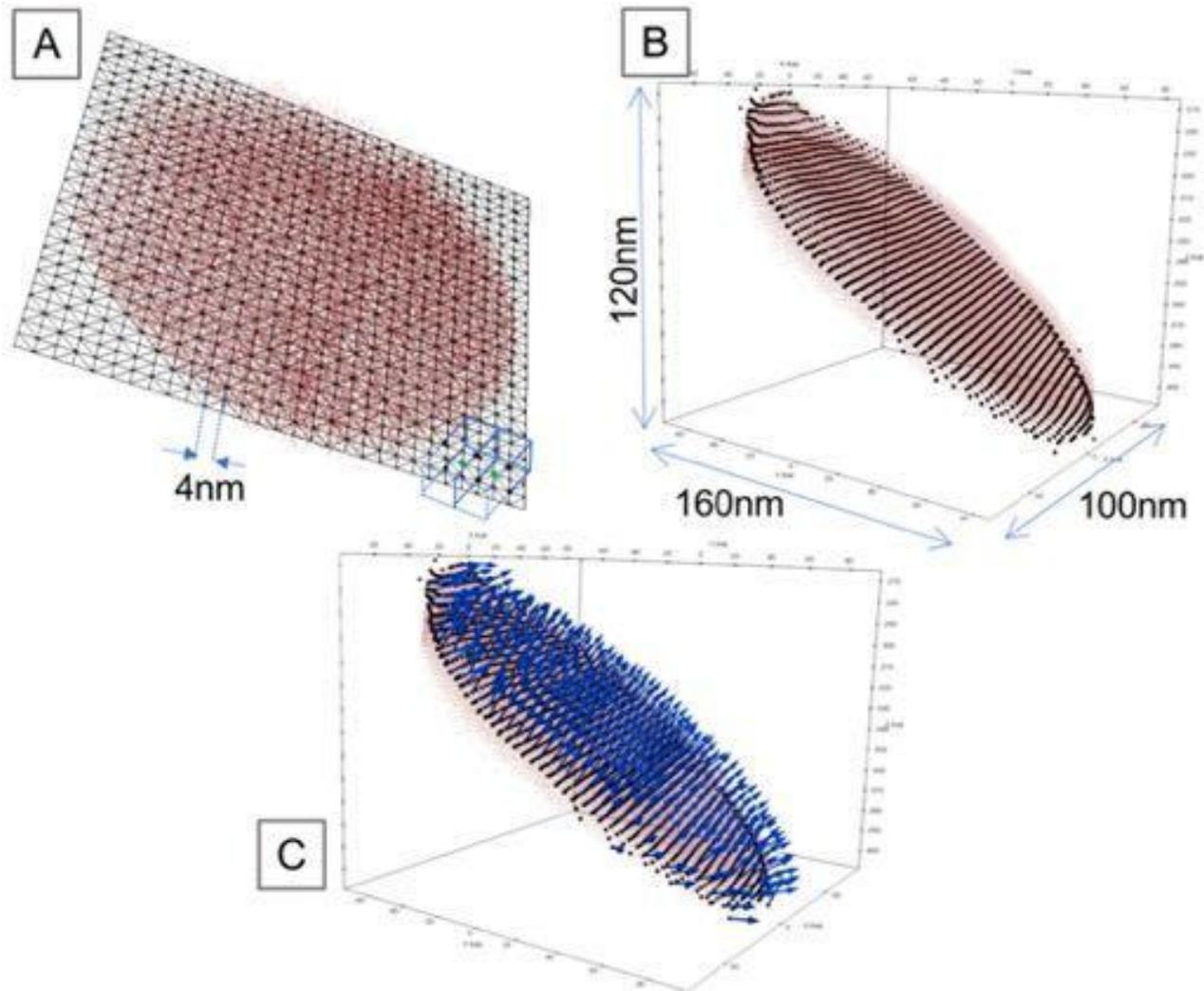

**Fig. 13. (A)** Voxel-centroids that comprise the plate-like precipitate are transformed into PCA coordinate system. A 2D triangular grid is superimposed on the precipitate in the plane formed by the PCA vectors with maximum standard deviation. The grid triangles not containing voxel centroids are removed. **(B)** Each node of the grid is moved to the center of mass of the voxel-centroids contained in the cuboid region centered at a given node. The cuboidal region for two green nodes is shown in panel **(A)**. As a result the grid takes the shape of the precipitate. **(C)** At each node the normal to the grid surface is evaluated and atoms present in a cylindrical region of interest (ROI) along the normal direction are selected. Further, average composition, thickness and 1D composition profiles are calculated along each ROI which enables the investigation of in-plane composition and thickness fluctuations.

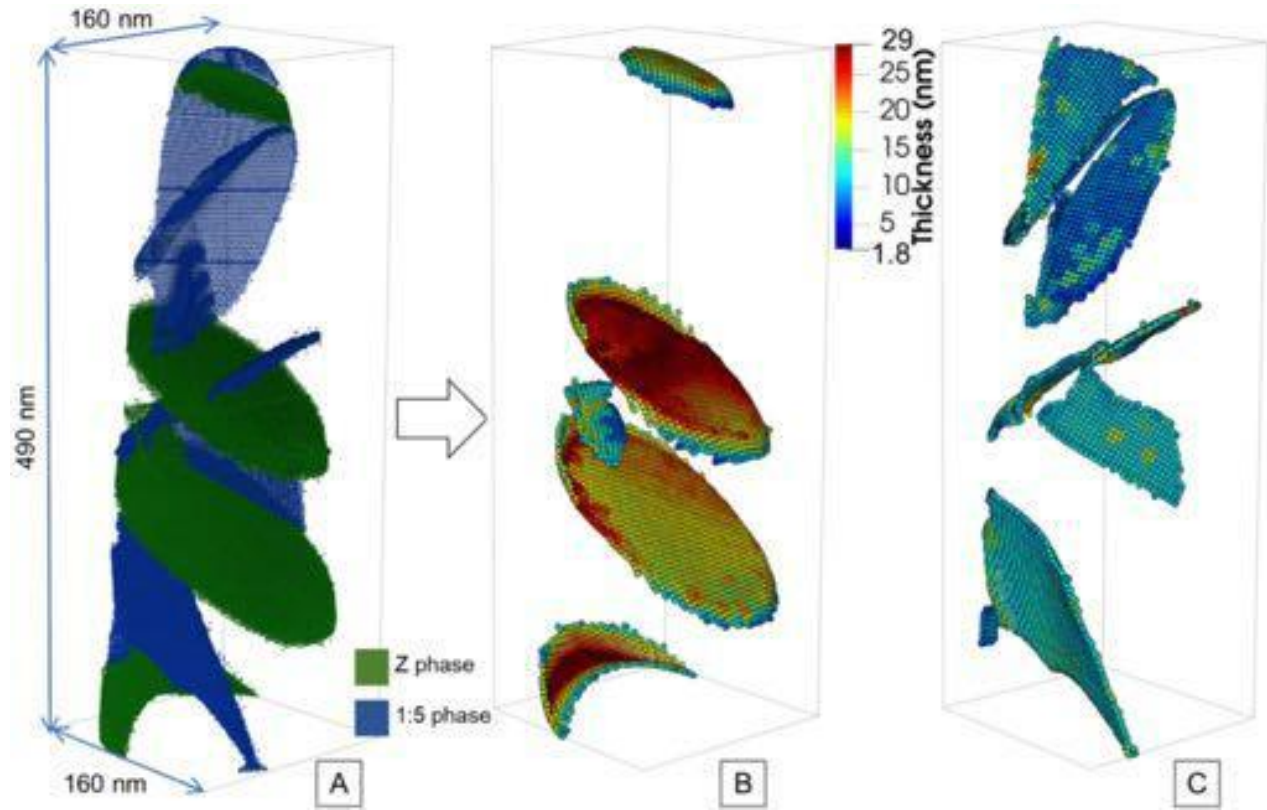

**Fig. 14. (A)** Voxel-centroids of the Z (green) and 1:5 (blue) phase are shown. Nodes of the triangular grid fitted on each plate like precipitate with thickness (nm) of the precipitate at each node is illustrated in **(B)** and **(C)**